\documentclass[12pt,preprint]{aastex}
\usepackage{amsmath,amssymb}

\usepackage{emulateapj5,apjfonts}

\journalinfo{The Astrophysical Journal, 576:806-813, 2002 September 10 
({\rm e-print} {\tt astro-ph/0203219})}

\def\pseudofigureone#1#2#3{{
\refstepcounter{figure}
\label{#1}
\vskip2mm
\plotone{#2}
\footnotesize\def\baselinestretch{1.0}
\begin{minipage}{\columnwidth}
{\scshape ~~Fig.}\space\thefigure.--- #3
\end{minipage}
\vskip2mm
}}
\def\pseudofiguretwo#1#2#3#4{{
\begin{figure*}
\epsscale{1.4}
\plottwo{#2}{#3}
\epsscale{1.0}
\caption{#4 
\label{#1}}
\end{figure*}
}}

\slugcomment{Received 2002 March 28; accepted 2002 May 15}
\shorttitle{SMALL-SCALE STRUCTURE OF MHD TURBULENCE}
\shortauthors{SCHEKOCHIHIN ET AL.}

\def\ssecref#1{\hbox{\S\,\ref{#1}}}

\def\secref#1{Sec.~\ref{#1}}

\def\exref#1{(\ref{#1})}

\def\eqref#1{Eq.~(\ref{#1})}
\def\eqsref#1{Eqs.~(\ref{#1})}

\def\figref#1{Fig.~\ref{#1}}

\def\const{{\rm const}}

\def\bea{\begin{eqnarray}}
\def\eea{\end{eqnarray}}

\def\phi{\varphi}

\def\({\left(}
\def\){\right)}
\def\[{\left[}
\def\]{\right]}
\def\<{\left\langle}
\def\>{\right\rangle}

\def\bl{\bigl}
\def\br{\bigr}
\def\la{\langle}
\def\ra{\rangle}

\def\d{\partial}
\def\diff{d}
\def\Dt{{d\over dt}\,}

\def\unity{{\hat{\mathbb I}}}

\def\vx{{\bf x}}
\def\vy{{\bf y}}

\def\vu{{\bf u}}
\def\vB{{\bf B}}

\def\vF{{\bf F}}
\def\vf{{\bf f}}
\def\vK{{\bf K}}
\def\vb{{\skew{-4}\hat{\bf b}}}
\def\vn{{\hat {\bf n}}}

\def\um{u_{\rm m}}
\def\vuext{\vu_{\rm ext}}
\def\vum{\vu_{\rm m}}
\def\Brms{B_{\rm rms}}
\def\Btyp{B_{\rm typical}}

\def\kperp{k_\perp}
\def\kpar{k_{\parallel}}

\def\lpar{\ell_{\parallel}}

\def\kd{k_{\nu}}

\def\kres{k_{\eta}}

\def\Re{{\rm Re}}

\def\Pr{{\rm Pr}_{\rm m}}

\def\Bf{B_{\rm f}}
\def\Kf{K_{\rm f}}
\def\Bb{B_{\rm b}}
\def\Kb{K_{\rm b}}
\def\lpar{l_\parallel}
\def\lperp{l_\perp}
\def\lb{l_{\rm b}}

\def\cK{{\alpha}}

\def\Bsq{{\langle B^2 \rangle}}

\def\Bfr{{\langle B^4 \rangle}}

\def\Ksq{{\langle K^2 \rangle}}
\def\Fsq{{\langle F^2 \rangle}}

\begin{document}

\title{THE SMALL-SCALE STRUCTURE OF MAGNETOHYDRODYNAMIC TURBULENCE WITH 
LARGE MAGNETIC PRANDTL NUMBERS}
\author{Alexander~A.~Schekochihin,
Jason~L.~Maron, 
Steven~C.~Cowley}
\affil{Plasma Physics Group, Imperial College, 
Blackett Laboratory, Prince Consort Rd., 
London~SW7~2BW, England}
\email{a.schekochihin@ic.ac.uk, maron@tapir.caltech.edu, 
steve.cowley@ic.ac.uk}
\author{James~C.~McWilliams}
\affil{Department of Atmospheric Sciences, 
UCLA, Los Angeles, California 90095-1565}
\email{jcm@atmos.ucla.edu}

\begin{abstract}

We study the intermittency and field-line structure of 
the MHD turbulence in plasmas with very large magnetic Prandtl 
numbers. In this regime, which is realized in the interstellar 
medium, some accretion disks, protogalaxies, 
galaxy-cluster gas, early Universe, etc., 
magnetic fluctuations 
can be excited at scales below the viscous cutoff. The salient 
feature of the resulting small-scale magnetic 
turbulence is the folded structure of the fields. 
It is characterized by very rapid transverse spatial 
oscillation of the field direction, while 
the field lines remain largely unbent up to the scale of the flow. 
Quantitatively, the fluctuation level and the field-line geometry 
can be studied in terms of the statistics 
of the field strength and of the field-line curvature. 
In the kinematic limit, 
the distribution of the field strength is an expanding 
lognormal, while that of the field-line curvature~$K$ is stationary 
and has a power tail~$\sim K^{-13/7}$. 
The field strength and curvature are anticorrelated,
i.e. the growing fields are mostly flat, while the 
sharply curved fields remain relatively weak. 
The field, therefore, settles into a reduced-tension state.
Numerical simulations demonstrate three essential features of 
the nonlinear regime. First, the total magnetic energy is 
equal to the total kinetic energy. Second, the intermittency is 
partially suppressed compared to the kinematic case, 
as the fields become more volume-filling and their 
distribution develops an exponential tail. 
Third, the folding structure 
of the field is unchanged from the kinematic case: 
the anticorrelation between the field strength and 
the curvature persists and the distribution of the latter retains
the same power tail. We propose a model of back reaction 
based on the folding picture that reproduces all of the above 
numerical~results.  

\end{abstract}



\section{INTRODUCTION}

In turbulent 
MHD systems where the ratio of fluid viscosity and magnetic diffusivity 
(the magnetic Prandtl number,~$\Pr=\nu/\eta$) is very large, 
there exists a broad range of subviscous scales available 
to magnetic fluctuations, but not to hydrodynamic ones.
This MHD regime is encountered, for example, in such astrophysical 
environments as the interstellar medium and 
protogalactic plasmas, where $\Pr$ can be as large 
as $10^{14}$ to $10^{22}$ \citep{Kulsrud_review}. 
Since the ratio 
of the resistive and viscous cut-off wave numbers is 
$k_\eta/k_\nu\sim\Pr^{1/2}$, 
this gives rise to subviscous scale ranges 7 to 11 decades wide.

Since the fluid is highly conducting, the magnetic-field lines 
are (nearly) perfectly frozen into the fluid flow. 
The fluid motions, even though restricted to the scales above the 
viscous cutoff, can excite magnetic fluctuations at much smaller 
scales via stretching and folding of the field lines. 
This possibility was first indicated by \citet{Batchelor}. 
The weak-field (kinematic) limit has been an attractive object 
of analytical study since the seminal work of \citet{Kazantsev}. 
The spectral theory of the kinematic dynamo driven by a random 
velocity field predicts exponential growth of the magnetic energy 
and its accumulation at the resistive scales \citep[see][and 
references therein]{Kazantsev,KA,Gruzinov_Cowley_Sudan,SBK_review}. 
More recently, it was realized that small-scale magnetic fields 
generated by this ``stretch-and-fold'' dynamo possess a 
distinctive spatial {\em folding structure} (\figref{fig_folded_lines}): 
the smallness of the field scale is due to rapid transverse 
spatial oscillation of the field direction, while the field lines 
remain largely unbent up to the scale of the 
flow~\citep[][--- this last paper is henceforth referred to 
as SCMM02]{Ott_review,Kinney_etal,SCMM_folding}. 

With the dramatic increase in the reach of the numerical experiment, 
the nonlinear regime became increasingly 
amenable to detailed study. The pioneering work of 
\citet{Meneguzzi_Frisch_Pouquet} was in recent years 
followed by a number of numerical investigations 
\citep[][the latter paper henceforth abbreviated MCM02]{Cattaneo_Hughes_Kim,Brandenburg_etal,Zienicke_Politano_Pouquet,Kinney_etal,Cho_Vishniac,Brandenburg,Chou,Brummell_Cattaneo_Tobias,MCM_dynamo}.
However, the physical difference between the $\Pr=1$ and $\Pr\gg1$ 
regimes is not always realized. 

Another defining physical feature of the MHD regime we are considering 
is the absence of an externally imposed uniform magnetic field. 
The difference is essential. First, a fixed uniform 
field implies a nonzero net flux through the system. Second, 
in the presence of a strong such field, magnetic-field lines cannot 
be bent, so the physics of the subviscous-scale magnetic 
fluctuations is more akin to that of the scalar turbulence 
\citep[cf.][]{Cho_Lazarian_Vishniac}.

In the astrophysical context, the main question has been of the impact 
small-scale magnetic fluctuations have on the feasibility of 
generating the large-scale galactic magnetic field by means of turbulent 
dynamo. In particular, one wonders how the accumulated small-scale 
magnetic energy affects the applicability of the mean-field 
$\alpha\Omega$-dynamo theory, which, in one form or another, 
has been at the center of all attempts to build a theoretical 
understanding of the galactic magnetic fields \citep{Beck_etal}. 
This issue remains unresolved in both the theoretical and numerical 
senses of the word. We note that, because of the large scale separations 
that have to be captured, the brute-force numerical solution 
of this problem remains beyond the capacity of currently available 
computational resources. On the other hand, the physics of the 
small-scale magnetic turbulence can be effectively studied 
as a separate problem and accessed (if only just) by 
numerical experiment. 
We hope that by developing a thorough physical understanding 
of the small-scale magnetic fluctuations, we can approach the 
problem of their interplay with the large-scale fields and motions. 

In this work, we concentrate on the structural properties 
of the small-scale magnetic fields: namely, the intermittency 
of their spatial distribution and the geometry of the field lines. 
Quantitatively, these are studied in terms of the one-point statistics 
of the field strength and of the field-line curvature. 

We consider the nonlinear turbulent dynamo to be described by the 
equations of incompressible~MHD:
\bea
\label{NSEq}
\Dt\vu &=& \nu\Delta\vu - \nabla p + \vB\cdot\nabla\vB + \vf,\\
\label{ind_eq}
\Dt\vB &=& \vB\cdot\nabla\vu + \eta\Delta\vB,
\eea
where $d/dt=\d_t+\vu\cdot\nabla$ is the convective derivative,
$\vu(t,\vx)$~is the velocity field, 
$\vB(t,\vx)$~is the magnetic field, 
and $\vf(t,\vx)$~is a random (white in time) large-scale forcing. 
The density~$\rho$ of the plasma is taken 
to be constant. The incompressibility condition $\nabla\cdot\vu = 0$ 
is, therefore, added to the equations above and serves to determine 
(or, indeed, to define) pressure. 
For the sake of convenience, the pressure~$p$ 
and the magnetic field~$\vB$ have been normalized 
to~$\rho$ and~$(4\pi\rho)^{1/2}$, respectively.

The plan of further proceedings is as follows.
In \secref{sec_kinematic}, we review the necessary 
facts about the kinematic regime of the dynamo. These are important, 
for they form the basis of our understanding of the dynamo 
and remain surprisingly relevant in the nonlinear regime. 
The latter constitutes our main object of study. 
In \secref{sec_nonlinear}, we describe the results of our 
numerical experiments and propose a heuristic physical model 
that explains these results. Conclusions are drawn 
in \secref{sec_conclusions}. 

\pseudofigureone{fig_folded_lines}{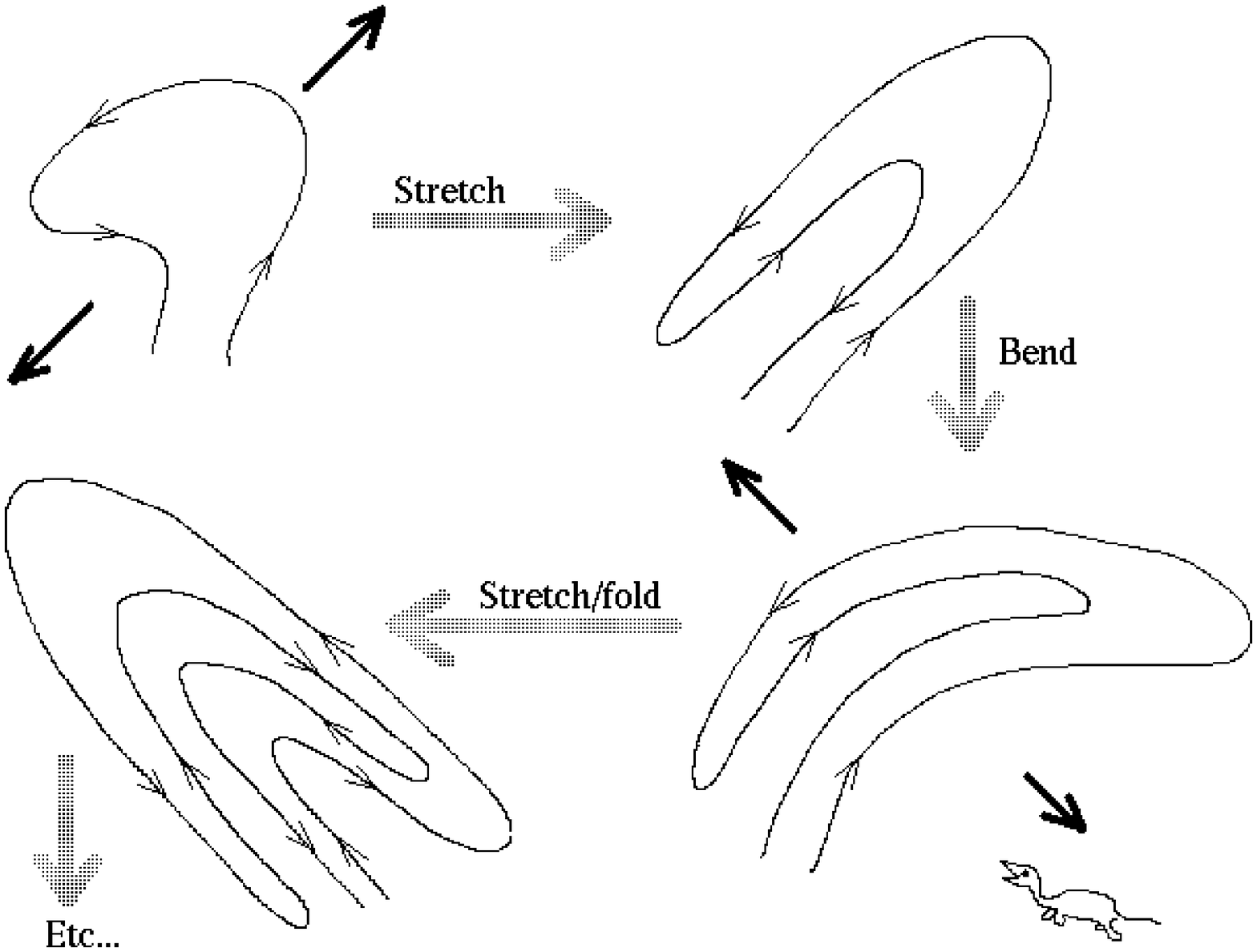}{Folding-structure 
formation via stretching of the field lines. 
Bold arrows indicate directions in which volumes are 
stretched by random shear.}

\section{THE KINEMATIC REGIME}
\label{sec_kinematic}

In the weak-field (kinematic) limit, 
MHD turbulence reduces to the problem of passive advection 
of a vector field by a turbulent velocity field. 
The magnetic energy grows exponentially and the small-scale 
folding structure is formed at the time scale associated with 
the eddies that turn over the fastest (i.e., in Kolmogorov 
turbulence, the viscous-scale eddies). 
Physically, this follows 
from the fact that the turbulent eddies act on 
the small-scale fields as a sequence of random 
linear-shear transformations (see \figref{fig_folded_lines} 
and~\ssecref{ssec_structure}). 
An expanding {\em lognormal} distribution of the field strength  
emerges, which is qualitatively 
explained in terms of the Central Limit Theorem. 

\subsection{The Kazantsev--Kraichnan Model}
\label{ssec_KKmodel}

All of the above results can be derived analytically 
in the framework of the \citet{Kraichnan_ensemble} 
model of passive advection,
which replaces the turbulent velocity with a Gaussian random field 
$\delta$-correlated in time:
\bea
\label{KK_field}
\<u^i(t,\vx)u^j(t',\vx')\> = \delta(t-t')\kappa^{ij}(\vx-\vx'). 
\eea
In the context of the small-scale 
dynamo, this model was first proposed by \citet{Kazantsev}.
While the Kazantsev--Kraichnan velocity 
field is, of course, highly artificial and does not 
approximate the real turbulent velocity field in any controlled 
sense, its performance in capturing the essential qualitative 
and, in some cases, also quantitative, features of the passive 
advection has been very impressive 
(on the passive scalar, 
see recent review by \citealt{Falkovich_Gawedzki_Vergassola}; 
on the kinematic dynamo, 
see \citealt{Kinney_etal}, SCMM02, MCM02, \citealt{SMCM_stokes}). 
This seems to suggest that 
the statistics of passive advection may be largely universal 
with respect to the structure of the ambient random flow. 

In the limit of large~$\Pr$, the magnetic fluctuations  
are mostly excited deep in the 
subviscous range, in which the fluid motions are strongly 
damped by viscous dissipation. The velocity field ``seen'' 
by the magnetic field is, therefore, regular and, in fact, 
effectively constitutes a {\em single-scale flow}. 
Most of the relevant statistical results turn out to depend 
just on the first few coefficients of the Taylor expansion 
of the velocity correlation tensor:
\bea
\nonumber
\kappa^{ij}(\vy) &=& \kappa_0\delta^{ij} 
- {1\over 2}\,\kappa_2\(y^2\delta^{ij} - {1\over2}\,y^iy^j\)\\ 
&& + {1\over 4}\,\kappa_4 y^2\(y^2\delta^{ij} - {2\over3}\,y^iy^j\) 
+ \cdots
\label{ss_exp}
\eea
(some of the coefficients in the expansion are fixed by 
the incompressibility constraint). Physically, 
$\kappa_2\sim\kd u$ is the stretching rate, while 
$\kappa_4$~sets the scale of the flow ($\kappa_4/\kappa_2\sim\kd^2$)
and is responsible for bending the field lines. 

\subsection{Intermittency}
\label{ssec_intermittency}

One of the simplest and most basic features of the kinematic 
diffusion-free regime is 
the lognormality of the distribution of the magnetic-field 
strength \citep[see, e.g.,][SCMM02]{BS_metric}. 
On the most fundamental level, the lognormal 
character of the distribution of~$B$ follows already from the form 
of the induction equation~\exref{ind_eq} (without the diffusion 
term). Indeed, it is linear with respect to the magnetic field, 
which is multiplied by a random externally prescribed 
velocity-gradient matrix~$\nabla\vu$. The Gaussianity of~$\log B$ 
then follows by the Central Limit Theorem. 

Within the framework of the Kazantsev--Kraichnan model, 
this result can be derived exactly. 
The one-point probability-density function 
(PDF) of the magnetic-field strength 
in the diffusion-free regime can easily be shown to satisfy 
the following Fokker--Planck equation:
\bea
\label{FPEq_B}
\d_t P = {1\over4}\,\kappa_2\,{\d\over\d B} B^4 
{\d\over\d B}\,{1\over B^2}\,P.
\eea
\eqref{FPEq_B}~written in log variables is simply 
a 1D~diffusion equation with a drift and has the following 
Green's-function solution: 
\bea
\nonumber
P(t,B) &=& {e^{-(1/2)\kappa_2 t}\over\sqrt{\pi\kappa_2 t}}
\int_0^\infty {\diff B'\over B'}\,P_0(B')\\
&&\times
\exp\(-{\bl[\ln(B/B') - (1/4)\kappa_2 t\br]^2\over\kappa_2 t}\),
\eea 
where $P_0(B)$ is the initial distribution. 
The lognormality means that the~PDF develops considerably 
spread-out tails at both large and small values 
of~$B$, i.e., the fluctuating magnetic fields in the kinematic regime 
possess a high degree of intermittency. The moments~$\la B^n\ra$ 
of the magnetic field have growth rates that 
increase quadratically with~$n$: 
$\la B^n\ra\propto\exp\bl[n(n+3)\kappa_2 t/4\br]$. 
The exponential growth of the moments of~$B$ is the manifestation 
of the dynamo action of the turbulence. It measures the effect 
of stretching of the field lines by the ambient random flow. 

Note that these results 
apply to the diffusion-free situation (i.e., to an ideally conducting 
medium), and, therefore, hold during the time it takes the magnetic 
excitation to propagate from the velocity scales, where it is 
assumed to be initially concentrated, to the resistive scales.  
This process occurs exponentially fast~\citep{KA,SBK_review}, 
so the corresponding time can be estimated 
as~$t\sim\kappa_2^{-1}\log\Pr^{1/2}$, 
which amounts to several eddy-turnover times. 
The PDF of~$B$ with account taken of diffusion has not as yet been found, 
though \citet{Chertkov_etal_dynamo} did, under certain additional 
assumptions, derive the moments~$\la B^n\ra$ in the diffusive regime. 
While the specific expressions for the growth rates 
of the moments are substantially modified, they still increase 
as~$n^2$, as would be the case for a lognormal distribution, 
so the intermittency is not diminished.

\subsection{The Folding Structure}
\label{ssec_structure}

The geometry of the magnetic-field lines can be studied in terms 
of the statistics of their curvature $\vK=\vb\cdot\nabla\vb$ 
(here~$\vb=\vB/B$). The evolution equation for the curvature 
can be derived from the induction equation: 
\bea
\nonumber
\Dt\vK &=& \vK\cdot\(\nabla\vu\)\cdot\bl(\unity-\vb\vb\br) 
- \vb\vK\vb:\nabla\vu - 2\vK\vb\vb:\nabla\vu\\
&& +~\vb\vb:\(\nabla\nabla\vu\)\cdot\bl(\unity - \vb\vb\br),
\label{K_eq}
\eea
where we have dropped diffusion terms. 
For the Kazantsev--Kraichnan velocity, 
the PDF of curvature satisfies the following Fokker--Planck 
equation (SCMM02)
\bea
\label{FPEq_K}
\d_t P = {7\over4}\,\kappa_2\,{\d\over\d K}\,K
\[\(1+K^2\){\d\over\d K}\,{1\over K}\,P + {20\over 7}\,P\],
\eea
where $K$ has been rescaled by $K_*=(12\kappa_4/7\kappa_2)^{1/2}\sim\kd$. 
From this equation one immediately finds that the curvature 
attains a stationary distribution:
\bea
\label{PK_kin}
P(K) = {7\over5}\,{K\over \(1 + K^2\)^{10/7}}.
\eea
This PDF has a power tail~$\sim K^{-13/7}$, which is in very good 
agreement with numerics (SCMM02). 
Naturally, if the resistive regularization 
is introduced, the power tail is cut off at 
$K\sim\kres\sim\Pr^{1/2}\kd$.
The bulk of the curvature distribution is concentrated at values 
of curvature~$K\sim\kd$, which reflects the prevailing 
straightness of the field lines at subviscous scales. 
The field lines are significantly curved only in the bends of the folds. 
While these bends occupy a small fraction of the volume, 
there is a high degree of intermittency in the distrubution 
of the characteristic scales associated with them. 
This is reflected by the power tail of the curvature PDF. 

Furthermore, 
{\em the field strength and the curvature are anticorrelated,} 
i.e. the growing fields are mostly flat, while the 
curved fields in the bends of the folds remain relatively 
weak \citep[cf.][]{Drummond_Muench,Brandenburg_Procaccia_Segel}. 
It is not difficult to realize that the nature of this anticorrelation, 
which is derived as a statistical property in~SCMM02, 
is, in fact, dynamical. Since this fact 
is important in our discussion of the nonlinear 
back reaction in~\ssecref{ssec_model}, we would like 
to explain it in more detail. 
 
\pseudofigureone{fig_one_fold}{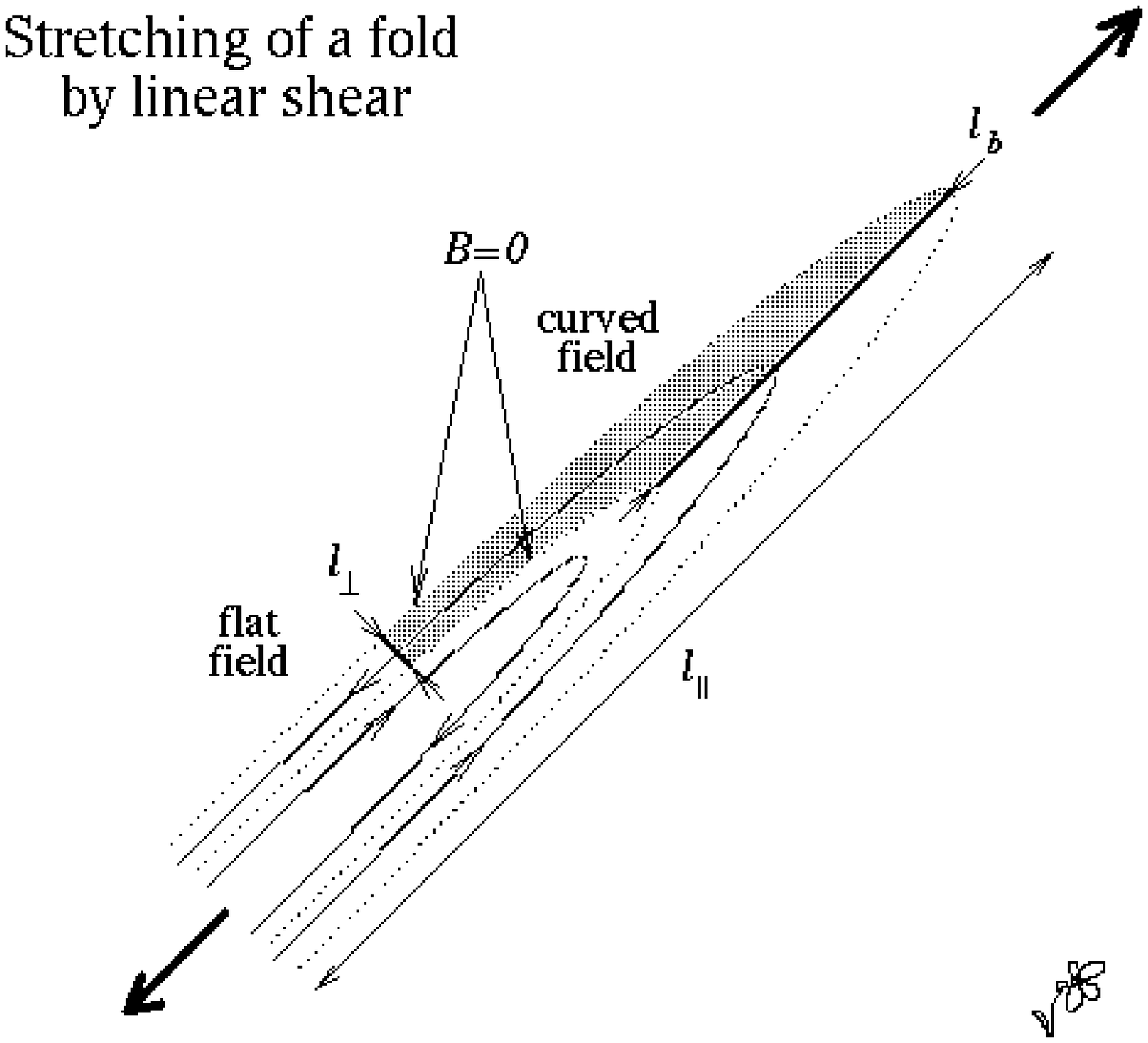}{Stretching of 
a fold by linear shear. The bold 
arrows indicate the stretching direction of the shear. 
The dotted lines correspond to surfaces where~$B=0$.} 

Let us examine what happens when a typical element of 
the folding structure (\figref{fig_one_fold}) is stretched 
by linear shear. Let $\Bf$ and $\Kf$ be the magnetic-field 
strength and curvature associated with the flat part of the 
fold, and $\Bb$ and $\Kb$ their counterparts in the 
curved part (the bend). In addition, let~$\lpar$ be the length of 
the fold (parallel scale of the field), $\lperp$~its 
thickness (perpendicular scale of the field), and 
$\lb$~the length of the bend.
\figref{fig_one_fold} can be thought of as a two-dimensional 
cross-section of a typical magnetic folding structure in 
three dimensions. It is assumed that the characteristic scale 
in the direction perpendicular to the page is 
approximately the same along the fold. 
Consider then the volume whose cross-section is indicated 
by the shaded area in~\figref{fig_one_fold}: 
all the flux is through the surfaces whose cross-sections 
are represented by bold lines.
Then, by flux conservation, we must have 
\bea
\label{flux_cons}
\Bf\lperp \sim \Bb\lb. 
\eea
Suppose that the fold is stretched by a factor~$s$ 
in the parallel direction. Then 
$\lpar\to s\lpar$, $\lb\to s\lb$, and, by volume and flux 
conservation, $\Bf\to s\Bf$, so 
the field is amplified in the flat part of the fold. 
We now observe that 
$\Kf\sim1/\lpar$ and $\Kb\sim1/\lperp$, whence 
$\Bf\Kf\sim\Bf/\lpar$ and, using~\eqref{flux_cons}, 
$\Bb\Kb\sim\Bb/\lperp\sim\Bf/\lb$. 
Both of these ratios remain unchanged during the stretching event. 
We conclude that linear shear transformation preserves 
\bea
\label{anticorr_dynamic}
KB \sim \const
\eea
throughout the fold: hence, the anticorrelation between $B$ and~$K$. 
Note that \eqref{anticorr_dynamic}~can be regarded as equivalent 
to the conjecture by \citet{Brandenburg_Procaccia_Segel} that flux 
tubes straighten at the same rate as the corresponding magnetic 
fields grow. 

\pseudofiguretwo{fig_BK_slices}{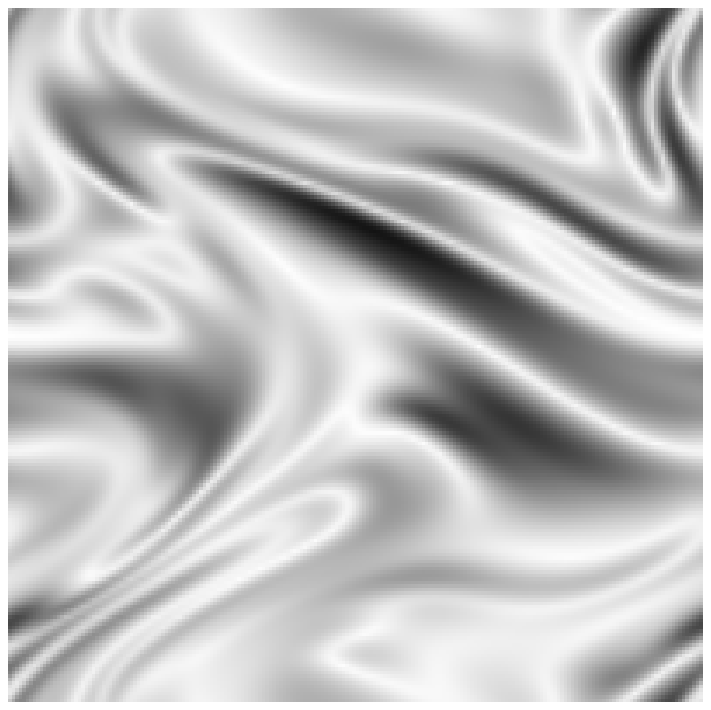}{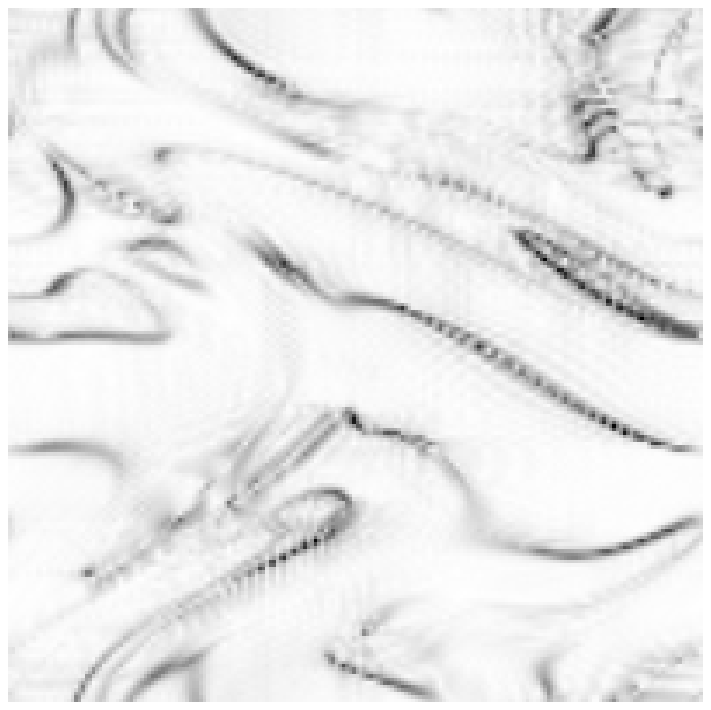}{Instantaneous cross-sections of the field strength~$B$ (left) 
and curvature~$K$ (right) during the nonlinear stage of 
a simulation with $\nu=5\cdot10^{-2}$, $\eta=2\cdot10^{-4}$ 
($\Pr=250$). Darker areas correspond to larger values. Detailed 
anticorrelation between $B$ and $K$ is manifest.} 

The perceptive reader might wonder about the consistency 
of~\eqref{anticorr_dynamic} with our previously stated results 
that, while the statistics of the field strength are lognormal, 
the curvature distribution has a power tail. The explanation is 
as follows. The argument that led to~\eqref{anticorr_dynamic} 
was based on approximating the ambient flow by a {\em linear} 
shear and thus described the effect of the first derivatives 
of the velocity on the magnetic field. We now observe that 
the curvature evolution equation~\exref{K_eq} also contains 
the second derivatives of velocity, which enter as a source term. 
Their role is to {\em bend} the field lines at the scale of the flow 
(see \figref{fig_folded_lines}): once the 
field becomes too flat (i.e., flatter than the flow), it is bent 
within one eddy-turnover time and $K\sim\kd$ is restored. 
Therefore, the curvature cannot remain below values of 
order~$\kd$ for a long time (or in many places). 
In formal terms, the second derivatives of 
velocity break the scale invariance of~\eqref{K_eq} and, 
consequently, of the associated Fokker--Planck 
equation~\exref{FPEq_K} (this is reflected in the presence 
of unity in the first term on the right-hand side).  
A stationary distribution with a power tail 
is thereby made possible. In fact, this power tail 
is a limiting envelope for a lognormal tail produced 
by the action of the linear shear (see SCMM02).


The most important consequence is that the field settles into a 
{\em reduced-tension state:} the tension force 
can be estimated by 
$\vB\cdot\nabla\vB\sim \kpar B^2 \sim KB^2 \sim\kd B^2$ on the average. 
A simple {\em reductio ad absurdum} argument can be envisioned 
to further support this statement. 
Let us write the evolution 
equation for~$\vF=\vB\cdot\nabla\vB$: 
\bea
\label{F_eq}
\Dt\vF = \vF\cdot\nabla\vu + \vB\vB:\nabla\nabla\vu,
\eea
where the diffusion terms are again dropped. Suppose 
for a moment that the field is chaotically tangled, 
i.e.~$F\sim\kpar B^2$ with~$\kpar\sim\kperp\gg\kd$. Then the term 
in~\eqref{F_eq} that involves the second derivatives of 
the velocity field can be neglected and \eqref{F_eq}~becomes 
formally identical to the evolution equation for~$\vB$. 
The moments of~$\vF$ must, therefore, grow at the same rates 
as the moments of~$\vB$, and we estimate 
$\Fsq/\Bfr\propto\Bsq/\Bfr$, 
which decays exponentially fast in time. 
An exact statistical calculation for 
the Kazantsev--Kraichnan velocity, shows that, indeed, 
\bea
{\la |\vB\cdot\nabla\vB|^2\ra\over\Bfr} \to 
{28\over9}\,{\kappa_4\over\kappa_2} \sim \kd^2 
\eea
asymptotically with time, starting from any initial conditions (SCMM02). 
The convergence is exponentially fast at the stretching (eddy-turnover) 
rate. We conclude that even an initially chaotically tangled magnetic 
field will quickly develop the folding structure.

Thus, the nonlinear saturation, which is due to the Lorentz tension 
balancing the stretching action of the flow, occurs when the energy 
of the field becomes comparable to the energy of the turbulent eddies. 
Note that in a hypothetical chaotically tangled field 
with $\kpar\sim \kperp$, 
the tension would be much larger: $\vB\cdot\nabla\vB\sim \kres B^2$, 
so saturation would be possible already at very low magnetic energies. 

\section{THE NONLINEAR REGIME}
\label{sec_nonlinear}

\subsection{Numerical Results}
\label{ssec_numerics}

No satisfactory 
analytical description of the nonlinear state is as yet 
available, so one must be guided by results of numerical 
experiments. The main obstacle in the way of a definitive 
numerical study is the tremendously wide range of scales that 
must be resolved in order to adequately simulate the large-Pr 
MHD: indeed, one must resolve {\em two} scaling intervals, 
the hydrodynamic inertial range and the subviscous magnetic one. 
Since this is not feasible, we propose to simulate 
the initial stage of the nonlinear evolution up to the 
point when the {\em total} energy of the magnetic field 
equalizes with the energy of the viscous-scale turbulent eddies. 
This stage can be studied in the {\em viscosity-dominated MHD regime} 
where the hydrodynamic Reynolds number~$\Re$ is order one 
and the external forcing models the energy supply from 
the larger eddies \citep[cf.][]{Cattaneo_Hughes_Kim,Kinney_etal}. 
Moreover, one can argue \citep[MCM02,][]{SCHM_ssim} 
that once the magnetic energy does equalize with 
that of the smallest eddies, the following scenario takes place. 
The magnetic back reaction leads to 
suppression of the shearing motions associated with 
the viscous-scale eddies. Larger-scale eddies, which are more 
energetic (but have slower turnover rates), continue 
to drive the small-scale magnetic fluctuations 
by the same stretching mechanism that the viscous-scale 
ones did, until 
the magnetic field becomes strong enough to suppress these 
eddies as well. This process continues until all field-stretching 
motions throughout the inertial range are 
suppressed \citep[see][for further discussion]{SCHM_ssim}. 
Speculatively, this could be thought of as some effective renormalization 
of Re, so that the final statistics of the magnetic field 
would again be described by the low-$\Re$ MHD model. Some numerical 
evidence supporting this picture is given by MCM02. 

In our simulations, we choose the parameters so that the forcing 
and the viscous scales are comparable and the statistics of 
the subviscous-range magnetic fluctuations can be studied 
already at~$128^3$ resolution. The numerical set-up and the spectral 
MHD code used are exhaustively described in MCM02. 
The forcing is large scale, nonhelical, and white in time.  
The units are based on the box size~1 and the forcing power~1. 
In these units, setting $\nu=5\cdot10^{-2}$ effectively leads 
to~$\Re\sim1$. 

After the initial kinematic growth stage, saturation is achieved 
where the total magnetic and hydrodynamic energies 
are equal.\footnote{See the argument at the end of \ssecref{ssec_structure}.
The equipartition only holds 
asymptotically with~$\Pr$. When the magnetic diffusivity is too large, 
saturation occurs at smaller magnetic energies 
\citep[for more details, see][]{SCHM_ssim,SMCM_stokes}. 
This probably accounts for subequipartition saturated 
states seen in some of the simulations of 
small-scale dynamo \citep[e.g.,][]{Brummell_Cattaneo_Tobias}. 
We note, however, that the lower magnetic-energy levels in 
the saturated state do not alter either the intermittency- 
or the structure-related properties of the small-scale fields.} 
We then measure the PDFs of the magnetic-field strength 
and curvature in the saturated state. Our findings are as follows.

\pseudofigureone{fig_PB_z281_hist}{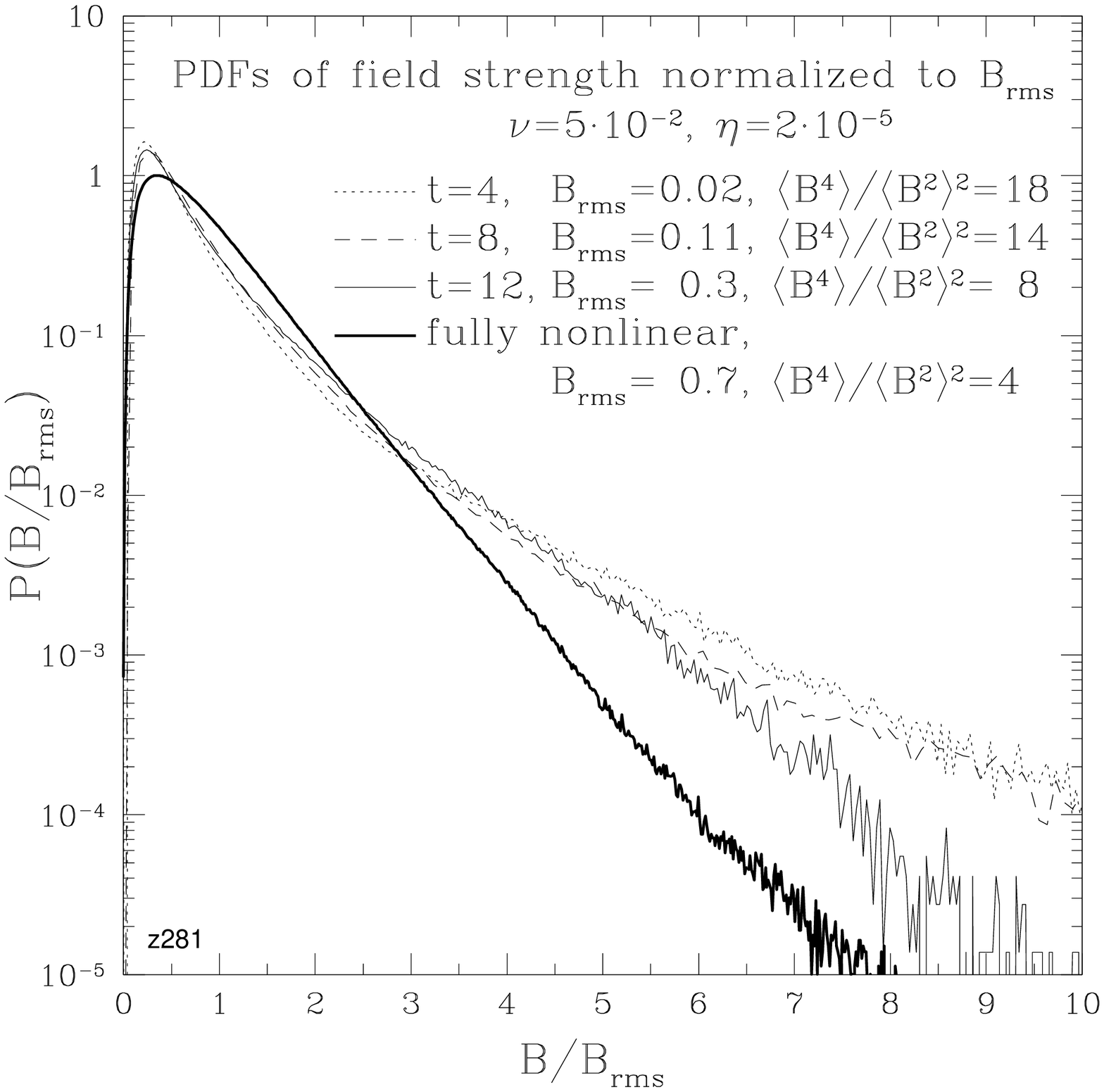}{PDFs of the field 
strength: onset of nonlinearity and saturation. 
The field strength is normalized to its rms value for each PDF. 
The intermittency at earlier times is thus clearly seen to be 
stronger than during the saturated stage.}

1.~{\em The level of intermittency is lower than in the kinematic case, 
the field-strength distribution developing an exponential tail.} 
The kurtosis of the field decreases to 
$\Bfr/\Bsq^2\sim 4$ 
after growing exponentially during the kinematic stage 
(for a Gaussian three-dimensional random field, the kurtosis 
would have been~$5/3$). 

The partial suppression of intermittency indicates a plausible 
scenario for the onset of nonlinearity 
(see \figref{fig_PB_z281_hist}). 
As we saw in~\ssecref{ssec_intermittency}, 
the kinematic dynamo gave rise to a highly intermittent 
lognormal spatial distribution of the field strength. 
Already early on, when the total magnetic energy is only 
a fraction of its saturated value, there are 
tiny regions where the magnetic-energy density 
locally approaches values comparable to, and greater than, 
the energy density of the fluid motions. 
This activates the nonlinear back reaction 
in these regions, so as to suppress further growth of the 
field there. As the overall amount of the magnetic energy grows, 
the fraction of the volume where the nonlinearity is at work 
increases until the nonlinear state is established globally. 
In the process, the magnetic fields 
become more volume-filling and, consequently, less 
intermittent \citep[cf.][]{Brandenburg_etal,Cattaneo_review,Cattaneo_solar}.  
The final PDF in the fully nonlinear regime has an exponential 
tail (\figref{fig_PB_z281_hist}): still considerably intermittent, 
but less so than the lognormal distribution. 

\pseudofigureone{fig_PK_eta}{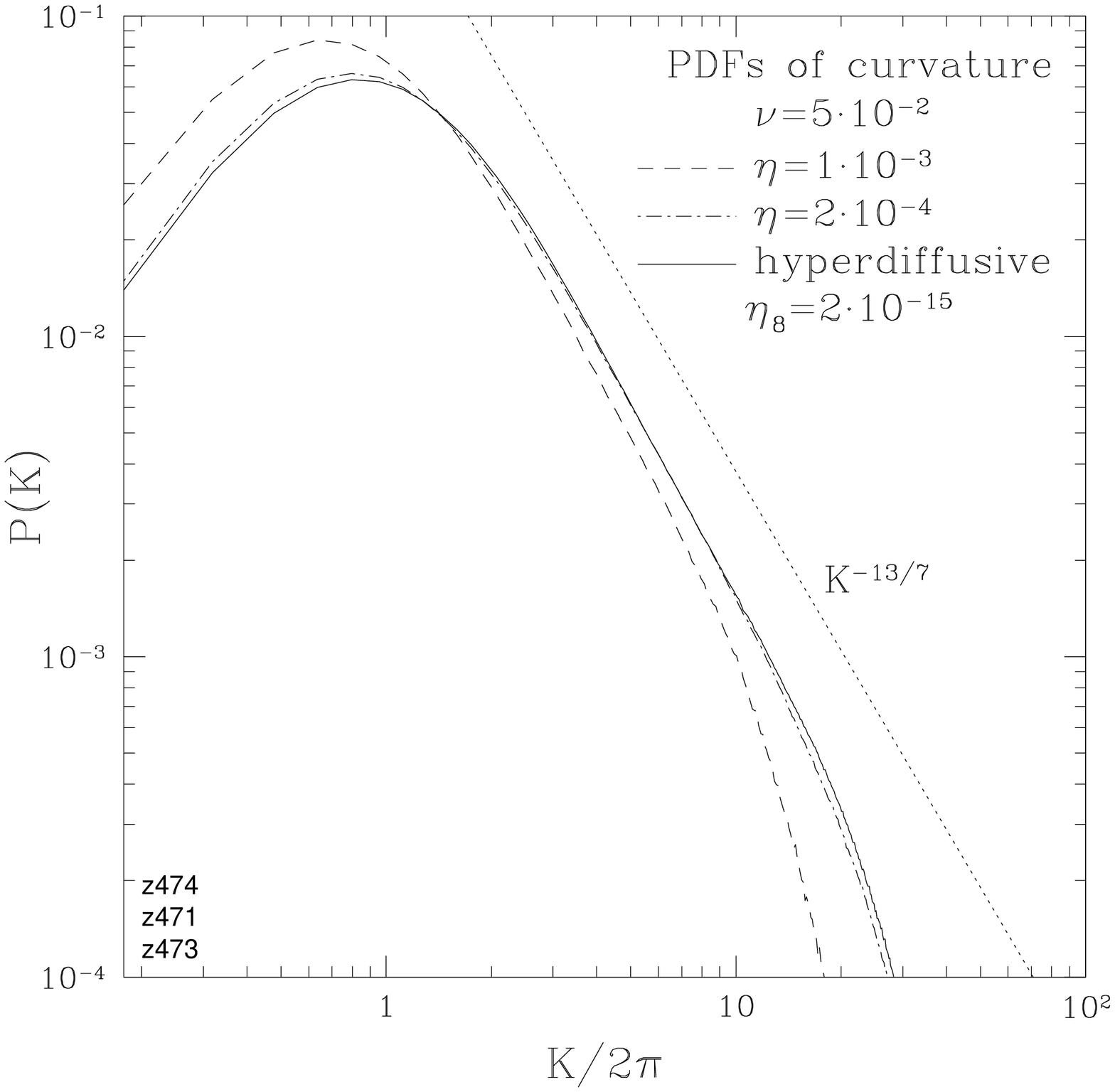}{PDFs of the field-line 
curvature in the nonlinear regime. The solid 
curve is for a simulation that used 8th-order hyperdiffusion. 
At curvatures approaching resistive cutoff, $K\sim\kres$, 
the scaling is destroyed.}

2.~{\em The folding structure of the field is unchanged from 
the kinematic case.} 
The anticorrelation between the field strength and 
the curvature persists (see \figref{fig_BK_slices}, which illustrates 
the dynamical nature of this anticorrelation discussed 
in~\ssecref{ssec_structure}). 
Their correlation coefficient 
$r_{K,B}=\la K^2 B^2\ra/\Ksq\Bsq - 1\sim -0.7$ in our simulations, 
which is quite close to its minimum allowed value of~$-1$.
Even more remarkably, 
the distribution of the curvature retains 
the same power tail~$\sim K^{-13/7}$ (\figref{fig_PK_eta}). 
We find this last feature particularly striking: 
indeed, not only does a kinematic theory based on the 
synthetic Kraichnan velocity field predict a quantitatively correct 
nontrivial scaling for the tail of the curvature PDF, but this scaling 
also survives in the fully nonlinear case! 


Finally, we would like to offer the following caveat with regard 
to the numerical study of the small-scale magnetic turbulence 
by means of spectral methods. 
We have found that if the amount 
of dissipation applied to the magnetic field is not sufficiently 
large to completely override the effect 
of dealiasing \citep[see][]{Canuto_etal}, 
the folding structure is destroyed and incorrect 
curvature statistics are obtained (cf.~MCM02). 
The resulting resolution 
constraints tend to be quite stringent: it is not enough 
to simply make sure that the UV cutoff of magnetic-energy spectrum 
is resolved. We believe that the destruction of the structural properties 
of the field by dealiasing is due to nonlocal nature of 
the corresponding operator. Consistently with this view, 
hyperdiffusion does not exhibit the same adverse effect 
(see \figref{fig_PK_eta}).

\subsection{A Model of Back Reaction}
\label{ssec_model}

The physical reason for the preservation of the folding structure 
is, of course, that the fluid motions at subviscous scales 
are strongly damped and cannot ``unwrap'' the folds. 
Here we use the folding picture to devise a simple {\em ad hoc} 
physical model that explains the two key quantitative 
results that have emerged from our numerics: the exponential tail 
of the PDF of the magnetic-field strength and the unchanged kinematic 
power tail of the curvature PDF.  

It is clear that, once the magnetic fields grow strong 
enough, they will resist further stretching by the ambient velocity 
shear. This increased rigidity of the field can be modelled by adding 
a nonlinear relaxation term to the induction 
equation: 
\bea
\label{ind_eq_rlx}
\Dt\vB = \vB\cdot\nabla\vuext - \tau_r^{-1}(B)\,\vB,
\eea
where $\vuext$~is the part of the velocity field due to 
the external forcing 
and $\tau_r(B)$~is some effective nonlinear relaxation 
time.\footnote{A model of nonlinear feedback mathematically very similar 
to this one was studied by \citet{ZMRS_nonlin} 
in the problem of self-excitation of a nonlinear 
scalar field in a random medium.} 

In view of the heuristic way in which 
the nonlinearity has been introduced into the induction equation, 
\eqref{ind_eq_rlx} {\em cannot} be expected to, and, indeed, 
does not, correctly model the spatial properties of the magnetic field.
What it effectively amounts to is a simple {\em local} one-point closure.
If the external velocity 
field is modelled by the Kazantsev--Kraichnan velocity, 
the Fokker--Planck equation~\exref{FPEq_B} derived in the kinematic case 
can easily be amended to include the nonlinearity:
\bea
\label{FPEq_nlin}
\d_t P = {\d\over\d B}\,B
\[{1\over4}\,\kappa_2 B^3{\d\over\d B}{1\over B^2}\,P + \tau_r^{-1}(B) P\].
\eea
This equation has one normalizable stationary solution 
\bea
\label{Pst_tau}
P(B) = \const\,B^2\exp\[-{4\over\kappa_2}
\int_0^B{\diff B'\over B'}\,\tau_r^{-1}(B')\].
\eea 

In order to make an intelligent guess about the $B$-dependence 
of~$\tau_r^{-1}(B)$, we have to invoke our 
understanding of the field structure at subviscous scales.
The relaxation term 
in~\eqref{ind_eq_rlx} is due to the suppression of the velocity 
shear by the Lorentz back reaction. 
The latter can be thought of as inducing a certain magnetic 
component~$\vum$ of the velocity field that counterbalances 
the external shear, so $\tau_r^{-1}(B)\sim\nabla\vum$. 
Since we are considering a viscosity-dominated 
regime, $\nabla\vum$~should be estimated by balancing 
the viscous damping and the magnetic tension: 
\bea
-\nu\Delta\vum \sim \vB\cdot\nabla\vB.
\eea 
The characteristic scale of~$\vum$ cannot be 
smaller than the viscous scale: otherwise, $\vum$~would be
viscously damped at a shorter time scale than that of the 
external shear and could not provide an 
effective counteraction to it. We have, therefore, 
\bea
\label{tau_general}
\tau_r^{-1}(B) \sim \kd\um \sim {\vB\cdot\nabla\vB\over\nu\kd} 
\sim {K B^2\over\nu\kd}.
\eea
We emphasize that the above relation represents a {\em local} 
force balance in a particular fold. 
Now, in a {\em typical} case, 
the folding structure would imply $K\sim\kd$, so 
\bea
\label{tau_typical}
\tau_r^{-1}(B) \sim {B^2\over\nu}, \qquad 
B\sim\Btyp, 
\eea
where $\Btyp\sim\Brms$ is the typical value of the 
back-reacting magnetic field in the system. 
This regime corresponds to what is sometimes referred to 
as the {\em the viscous relaxation effect} 
\citep{Chandran_visc_rlx,Kinney_etal}. 
Substituting the expression~\exref{tau_typical} 
for~$\tau_r^{-1}(B)$ into~\eqref{Pst_tau}, 
we find a Gaussian PDF \citep[cf.][]{Boldyrev_alpha} with 
$\Brms\sim(\kappa_2\nu)^{1/2}$, which approximately 
corresponds to magnetic energy equalizing with the energy 
of the viscous-scale eddies and accounts for the energy 
equipartition we have seen in our simulations. 

However, these arguments are only appropriate in application 
to the statistics of the {\em typical} values of the magnetic-field 
strength, i.e., to the {\em core} of the distribution 
function \citep[cf.][]{Boldyrev_alpha}. In order to understand 
the {\em tail} of the PDF, one must consider {\em rare} events, 
specifically, the instances (or places) where $B$ is 
{\em atypically large}. In the framework of the ``stretch-and-fold'' 
small-scale dynamo mechanism, such events can occur if 
magnetic field gets ``overstretched'' (for instance, because the 
shear has acted coherently in one direction for an atypically 
long time). The local relation between the curvature and the field 
strength associated with it is given by~\eqref{anticorr_dynamic}, 
whence $KB\sim\kd\Btyp$. 
Inserting this into~\eqref{tau_general}, we find 
\bea
\label{tau_tail}
\tau_r^{-1}(B) \sim {B\Btyp\over\nu}, \qquad 
B\gg\Btyp, 
\eea
which, upon substitution into~\eqref{Pst_tau}, gives 
the exponential tail evidenced by the numerics. 

The persistence of the kinematic curvature statistics 
in the face of nonlinear effects can also be understood in these 
terms. Essentially, the tail of the curvature PDF remains  
unaffected by the back reaction because it describes areas 
of anomalously large curvature ($K\gg\kd$) where, due to 
the anticorrelation property, the field is weak. 
On a slightly more quantitative level, we argue that 
the effect of the back reaction on the curvature can also 
be modelled by a simple nonlinear relaxation term, as 
in~\eqref{ind_eq_rlx}:
\bea
\label{K_eq_rlx}
\Dt\vK = \bl[{\rm rhs~of~\eqref{K_eq}~with}~\nabla\vuext\br] 
- \tau_r^{-1}(B)\,\vK. 
\eea
Here $\tau_r^{-1}(B)$~is again estimated via~\eqsref{tau_general} 
and~\exref{anticorr_dynamic}. 
The nonlinear relaxation term in~\eqref{K_eq_rlx} is then  
\bea
- \tau_r^{-1}(B)\,\vK \sim - {(KB)^2\over\nu\kd}\,\vn 
\sim -{\kd\Btyp^2\over\nu}\,\vn,
\eea
where~$\vn=\vK/K$. With this correction, 
the Fokker--Planck equation~\exref{FPEq_K} for 
the PDF of curvature becomes 
\bea
\d_t P = {7\over4}\,\kappa_2\,{\d\over\d K}\,K
\[\(1+K^2\){\d\over\d K}\,{1\over K}\,P + {20\over 7}\,P 
- {\cK\over K}\,P\],
\eea
where $\cK$~is a numerical constant of order unity and, as before, 
$K$~is rescaled by~$K_*\sim\kd$. It is a straightforward exercise 
to show that the stationary PDF is now given by
\bea
\label{PK_nlin}
P(K) = \const\,{K\,e^{\cK\arctan(K)}\over \(1 + K^2\)^{10/7}},
\eea 
which has the same power tail~$\sim K^{-13/7}$ as its kinematic 
counterpart~\exref{PK_kin}.

Finally, we would like to stress the qualitative character of 
the ideas and results put forward in this section. 
A more quantitative nonlinear theory based on these ideas may 
be feasible, but is left for future work. 
Another important issue that requires careful quantitative treatment 
is the role of Ohmic diffusion. Based on the model 
of back reaction proposed here, we seem to be able to understand 
the nonlinear regime without including the resistive terms. 
These terms are hard to treat analytically due to the usual 
closure problem associated with the diffusion operator. 
Numerically, we have confirmed that the field 
structure is unaffected by a switch to hyperdiffusion 
(see \figref{fig_PK_eta}), but the locality of 
the dissipation operator probably does matter 
(see remarks at the end of~\ssecref{ssec_numerics}). 
Further investigation of the universality of the statistics 
of the small-scale magnetic turbulence with respect to 
the form of the UV regularization is also left for the future.

\section{CONCLUSIONS} 
\label{sec_conclusions}

We have found that the folding structure of subviscous-scale 
magnetic fluctuations that is formed via the kinematic 
``stretch-and-fold'' small-scale-dynamo mechanism remains 
the essential feature of the nonlinear regime. 
The scale separation is of crucial importance here: 
while small-scale structure can be generated and maintained 
in large-scale random flows, these flows lack the ability 
to coherently undo the structure even when acting in concert 
with the magnetic back reaction. 

Both our theoretical arguments and our numerical experiments 
were based on viewing the turbulent velocity field as a single-scale 
random flow. In real turbulence, $\Re$~is, of course, fairly 
large ($10^4$~in the ISM), so many hydrodynamic scales 
come into play. Unfortunately, the resulting 
scale ranges are too broad to be adequately simulated. 
It is interesting, however, that the results presented above 
appear to hold in simulations with more realistic~Re (up to~$10^3$) 
but relatively small Pr (down to values of order~one), in 
accordance with the arguments presented at the beginning 
of~\secref{sec_nonlinear}. We believe, therefore, that 
the physics and the numerics we have laid out 
provide at least a semiquantitative description of 
the statistical properties of the small-scale MHD turbulence 
in high-$\Pr$ plasmas. 

The implications for astrophysical objects can be significant. 
For the large-scale galactic dynamo, small-scale fields must 
be taken into consideration if a nonlinear $\alpha\Omega$ theory 
is to be constructed. If the net effect of the accumulated 
small-scale magnetic energy is to suppress~$\alpha$, an alternative 
theory will have to be sought. 
The nonlinear evolution of the small-scale magnetic turbulence 
in protogalaxies \citep{Kulsrud_etal_proto} and in the 
early Universe \citep[see, e.g.,][]{Son,Christensson_Hindmarsh_Brandenburg} 
determines the energy and the coherence scale of the seed field 
inherited by newly formed galaxies. 
The structure of tangled magnetic fields in the intracluster 
gas crucially affects thermal conduction in the galaxy 
clusters \citep[see, e.g.,][]{Chandran_Cowley,Malyshkin_cond,Narayan_Medvedev}.
In the accretion-disk physics, presence of large amounts 
of small-scale magnetic energy could lead to new models 
for the angular-momentum transport \citep{Balbus_Hawley_review} 
and for the acceleration of jets \citep[see, e.g.,][]{Heinz_Begelman}. 
In the solar astrophysics, the possibility was recently raised 
that a substantial part of the magnetic energy in the quiet 
photosphere of the Sun resides in small-scale 
magnetic fluctuations \citep{Cattaneo_solar}. 
Indeed, it is natural to expect that, just as turbulence itself, 
small-scale random magnetic fields are ubiquitous in the Universe. 

Finally, constructing a self-consistent physical theory of 
the small-scale magnetic turbulence constitutes a fascinating 
task in its own right. Though fifty years have passed since 
\citet{Batchelor} took the first steps down this road, 
an inquisitive researcher will still find surprises at every 
turn, and it might well be short-sighted to claim that we are 
able to discern the contours of the final destination. 


\acknowledgements

It is a pleasure to thank E.~Blackman, S.~Boldyrev, W.~Dorland, 
P.~Goldreich, G.~Hammett, R.~Kulsrud, L.~Malyshkin, A.~Shukurov, 
and D.~Uzdensky for stimulating discussions. 
Our work was supported by 
the UKAEA Agreement No.~QS06992, 
the EPSRC Grant No.~GR/R55344/01, 
the NSF Grants~No.~AST~97-13241 and AST~00-98670, 
and the USDOE Grant~No.~DE-FG03-93ER54~224.
Simulations were run on the supercomputers of the 
National Center for Supercomputing Applications 
and of the UK Astrophysical Fluids Facility.


\begin{thebibliography}{}

\bibitem[Balbus \& Hawley(1998)]{Balbus_Hawley_review}
Balbus, S.~A. \& Hawley, J.~F. 1998, Rev. Mod. Phys., 70, 1

\bibitem[Batchelor(1950)]{Batchelor}
Batchelor, G.~K. 1950, Proc. Roy. Soc. London, Ser.~A, 201,~405


\bibitem[Beck et al.(1996)]{Beck_etal}
Beck, R., Brandenburg, A., Moss, D., Shukurov, A., \& Sokoloff, D. 1996,  
\araa, 34,~155

\bibitem[Boldyrev(2001)]{Boldyrev_alpha}
Boldyrev, S.~A. 2001, \apj, 562, 1081

\bibitem[Boldyrev \& Schekochihin(2000)]{BS_metric} 
Boldyrev, S.~A. \& Schekochihin, A.~A. 2000, \pre, 62,~545

\bibitem[Brandenburg(2001)]{Brandenburg}
Brandenburg, A. 2001, \apj, 550,~824

\bibitem[Brandenburg et al.(1996)]{Brandenburg_etal}
Brandenburg, A., Jennings, R.~L., Nordlund, A., Rieutord, M., 
Stein, R.~F., \& Tuominen, I. 1996, 
J.~Fluid Mech., 306, 325 

\bibitem[Brandenburg, Procaccia, \& Segel(1995)]{Brandenburg_Procaccia_Segel} 
Brandenburg, A., Procaccia, I., \& Segel, D. 1995, Phys. Plasmas, 2, 1148 

\bibitem[Brummell, Cattaneo, \& Tobias(2001)]{Brummell_Cattaneo_Tobias}
Brummell, N.~H., Cattaneo, F., \& Tobias, S.~M. 2001, 
Fluid Dyn. Res., 28, 237

\bibitem[Canuto et al.(1988)]{Canuto_etal}
Canuto, C., Hussaini, M.~Y., Quarteroni, A., \& Zang, T.~A. 1988, 
Spectral Methods in Fluid Dynamics (Berlin: Springer)

\bibitem[Cattaneo(1999a)]{Cattaneo_review}
Cattaneo, F. 1999a, 
in: Motions in the Solar Atmosphere, 
ed.~A.~Hanslmeier \& M.~Messerotti 
(Dordrecht: Kluwer), 119 

\bibitem[Cattaneo(1999b)]{Cattaneo_solar}
Cattaneo, F. 1999b, \apj, 515,~L39

\bibitem[Cattaneo, Hughes, \& Kim(1996)]{Cattaneo_Hughes_Kim}
Cattaneo, F., Hughes, D.~W., \& Kim, E. 1996, \prl, 76,~2057

\bibitem[Chandran(1998)]{Chandran_visc_rlx}
Chandran, B.~D.~G. 1998, \apj, 492,~179

\bibitem[Chandran \& Cowley(1998)]{Chandran_Cowley}
Chandran, B.~D.~G. \& Cowley, S.~C. 1998, \prl, 80, 3077

\bibitem[Chertkov et al.(1999)]{Chertkov_etal_dynamo}
Chertkov, M., Falkovich, G., Kolokolov, I., \& Vergassola, M. 1999, 
\prl, 83,~4065 

\bibitem[Cho, Lazarian, \& Vishniac(2002)]{Cho_Lazarian_Vishniac}
Cho, J., Lazarian, A., \& Vishniac, E.~T. 2002, \apj, 566, L49

\bibitem[Cho \& Vishniac(2000)]{Cho_Vishniac}
Cho, J. \& Vishniac, E.~T. 2000, \apj, 538, 217 

\bibitem[Chou(2001)]{Chou}
Chou, H. 2001, \apj, 556, 1038 

\bibitem[Christensson, Hindmarsh, \& Brandenburg(2001)]{Christensson_Hindmarsh_Brandenburg}
Christensson, M., Hindmarsh, M., \& Brandenburg, A. 2001, \pre, 64, 056405 

\bibitem[Drummond \& M\"unch(1991)]{Drummond_Muench}
Drummond, I.~T. \& M\"unch, W. 1991, J.~Fluid Mech., 225, 529 

\bibitem[Falkovich, Gaw\c{e}dzki, \& Vergassola(2001)]{Falkovich_Gawedzki_Vergassola}
Falkovich, G., Gaw\c{e}dzki, K., \& Vergassola, M. 2001, 
Rev.~Mod.~Phys, 73,~913 

\bibitem[Gruzinov, Cowley, \& Sudan(1996)]{Gruzinov_Cowley_Sudan}
Gruzinov, A., Cowley, S., \& Sudan, R. 1996, \prl, 77, 4342 

\bibitem[Heinz \& Begelman(2000)]{Heinz_Begelman}
Heinz, S. \& Begelman, M.~C. 2000, \apj, 535, 104

\bibitem[Kazantsev(1967)]{Kazantsev}
Kazantsev, A.~P. 1967, 
Zh. Eksp. Teor. Fiz., 53,~1806 
(Sov. Phys. JETP, 26,~1031)

\bibitem[Kinney et al.(2000)]{Kinney_etal}
Kinney, R.~M., Chandran, B., Cowley, S., \& McWilliams, J.~C. 2000, 
\apj, 545,~907 

\bibitem[Kraichnan(1968)]{Kraichnan_ensemble} 
Kraichnan, R.~H. 1968, Phys. Fluids, 11,~945

\bibitem[Kulsrud(1999)]{Kulsrud_review}
Kulsrud, R.~M. 1999, \araa, 37,~37

\bibitem[Kulsrud \& Anderson(1992)]{KA} 
Kulsrud, R.~M. \& Anderson, S.~W. 1992, \apj, 396,~606

\bibitem[Kulsrud et al.(1997)]{Kulsrud_etal_proto}
Kulsrud, R.~M., Cen, R., Ostriker, J.~P., \& Ryu, D. 1997, 
\apj, 480,~481


\bibitem[Malyshkin(2001)]{Malyshkin_cond}
Malyshkin, L. 2001, \apj, 554, 561 

\bibitem[Maron, Cowley, \& McWilliams(2002)]{MCM_dynamo}
Maron, J., Cowley, S., \& McWilliams, J. 2002, preprint astro-ph/0111008 
(MCM02)

\bibitem[Meneguzzi, Frisch, \& Pouquet(1981)]{Meneguzzi_Frisch_Pouquet}
Meneguzzi, M., Frisch, U., \& Pouquet, A. 1981, \prl, 47,~1060

\bibitem[Narayan \& Medvedev(2001)]{Narayan_Medvedev}
Narayan, R. \& Medvedev, M.~V. 2001, \apj, 562, L129

\bibitem[Ott(1998)]{Ott_review}
Ott, E. 1998, 
Phys. Plasmas, 5, 1636

\bibitem[Schekochihin, Boldyrev, \& Kulsrud(2002a)]{SBK_review}
Schekochihin, A.~A., Boldyrev, S.~A., \& Kulsrud, R.~M. 2002a, 
\apj, 567, 828

\bibitem[Schekochihin et al.(2002b)]{SCHM_ssim}
Schekochihin, A.~A., Cowley, S.~C., Hammett, G.~W., Maron, J.~L., 
\& McWilliams, J.~C. 2002b, 
``A model of nonlinear evolution and saturation of 
the turbulent MHD dynamo,'' New J.~Phys. (www.njp.org), submitted 

\bibitem[Schekochihin et al.(2002c)]{SCMM_folding}
Schekochihin, A., Cowley, S., Maron, J., \& Malyshkin, L. 2002c, 
\pre, 65, 016305 (SCMM02)

\bibitem[Schekochihin et al.(2002d)]{SMCM_stokes}
Schekochihin, A.~A., Maron, J.~L., Cowley, S.~C., \& McWilliams, J.~C. 2002d, 
``Simulations of the nonlinear small-scale dynamo,'' \apj, submitted

\bibitem[Son(1999)]{Son}
Son, D.~T. 1999, \prd, 59, 063008 



\bibitem[Zeldovich et al.(1987)]{ZMRS_nonlin}
Zeldovich, Ya.~B., Molchanov, S.~A., Ruzmaikin, A.~A., 
\& Sokoloff, D.~D. 1987, Proc.~Natl.~Acad.~Sci.~USA, 84,~6323

\bibitem[Zienicke, Politano, \& Pouquet(1998)]{Zienicke_Politano_Pouquet}
Zienicke, E., Politano, H., \& Pouquet, A. 1998, \prl, 81,~4640

\end{thebibliography}
\end{document}